\documentclass[aps,prl,showpacs,twocolumn,groupedaddress]{revtex4-1}

\usepackage{graphicx,graphics,color,epsfig}
\usepackage{amsmath}
\usepackage{amssymb}
\usepackage{amsfonts}
\usepackage{amsfonts}
\usepackage{epstopdf}
\usepackage{bm}
\usepackage{times,xspace}
\usepackage{color}

\begin{document}


\title{Discerning electronic fingerprints of nodal and antinodal nestings and their phase coherences in doped cuprate superconductors}

\author{Tanmoy Das}
\affiliation{Theoretical Division, Los Alamos National Laboratory, Los Alamos, New Mexico-87545, USA}

\date{November 25, 2012}
\begin{abstract}
The complexity of competing orders in cuprates has recently been multiplied by a number of bulk evidences of charge ordering with wavevector that connects the antinodal region of the Fermi surface. This results contradicts many spectroscopic results of the nodal nesting.
To resolve this issue, we carry out a unified study of the resulting electronic fingerprints of both nodal and antinodal nestings (NNs/ANs), and compare with angle-resolved photoemission, scanning tunneling spectroscopic data, as well as bulk sensitive Hall effect measurements. Our result makes several definitive distinctions between them in that while both nestings gap out the antinodal region, AN induces an additional quasiparticle gap {\em below the Fermi level along the nodal direction}, which is so far uncharted in spectroscopic data. Furthermore, we show that the Hall coefficient in the AN state obtains a discontinuous jump at the phase transition from an electron-like nodal pocket (negative value) to a large hole-like Fermi surface (positive value), in contrast to a continuous transition in the available data. Finally, we write down a Ginzburg-Landau functional to study the stability of the two phases, and discuss an interesting possibility of disorder pinned `chiral' charge ordering.
\end{abstract}

\pacs{74.72.Kf,74.25.Jb,74.25.F-,74.40.Kb}

\maketitle
Doped materials can accommodate multiform competing phases of matter, either in a uniform phase or phase separated,\cite{stripe} with a subclass of it that can inherit high-$T_c$ superconductivity. In cuprates, different theoretical routes to the mechanism of superconductivity are primarily motivated by the experimental evidences of different competing orders in the corresponding normal state. In particular, the well-established results of many bulk-sensitive probes have suggested a uniform or non-uniform nodal nesting (NN) which usually involves spin (and a possible interplay with charge excitations via incommensurability) modulations in La-based cuprates.\cite{LSCO} In stark contrast, recent measurements including scanning tunneling microscopy (STM)\cite{CDWSTM}, nuclear magnetic resonance (NMR) at finite magnetic field,\cite{CDWNMR} X-ray probes,\cite{CDWXray} and a thermodynamic measurement at high field,\cite{CDWtherm} indicate a charge modulation in Y-, Bi-based cuprates, arguably due to either uniaxial or biaxial antinodal nesting (AN). There also exist other possible experimental scenarios such as smectic\cite{smectic}, nematic,\cite{nematic} orbital loop orders,\cite{Greven} with various active degrees of freedom which can sometimes differ from spin and charge quanta. Therefore, discerning the correct nature of the competing phase, and their possible coexistence and competition is not only important to throw light on the pairing mechanism, but also to expand the choices of known emergent phases that can arise in an inhomogeneous environment.


From theoretical standpoint, the presently debated competing order scenarios of the pseudogap literature can mainly be classified into three categories: (1) A NN giving rise to Umklapp process,\cite{Rice} or $d$-density wave,\cite{DDW} or spin-ordering;\cite{Dastwogap} (2) An AN between Van-Hove singularity (VHS) region producing charge density wave (CDW);\cite{npocket} and (3) An incommensurate version of the NN involving both spin and charge excitations (`stripe'-phase).\cite{stripe,stripetheory} The perfect NN of any active order renders a nodal hole-pocket in hole-doped systems,\cite{Rice,DDW,Dastwogap} consistent with Luttinger volume counting. On the other hand, in recent works Harrison and co-workers,\cite{npocket} and Markiewicz {\it et al.}\cite{BobCDW} have demonstrated that the AN governs a nodal electron-pocket in these systems. Given that the shadow bands of the nodal pocket is difficult to detect unambiguously by angle-resolved photoemission spectroscopy (ARPES) and STM [via quasiparticle interference (QPI) technique], both scenarios can taken to be consistent with these data as long as only the Fermi surface (FS) topology is concerned. To resolve this issue, we carry out a mean-field calculation within single band model. A main conclusion of this Letter reveals that an electron-pocket in the nodal region leads several inconsistencies when compared to other spectroscopies. Since the nodal electron-pocket implies an additional quasiparticle gapping {\it along the nodal direction below the Fermi level ($E_F)$}, it leads  inconsistency when compared to well-established ARPES and STM results.\cite{arpes,QPI} The NN ${\bm Q}_n\sim(\pi,\pi)$, which yields nodal hole-pocket, and no nodal gap opening below $E_F$, is in detailed agreement with most features observed in spectroscopies. The `stripe' phase,\cite{stripetheory} creating many FS pockets in contrast to a single `Fermi arc', will not be discussed here.

To strengthen our conclusion, we also compute temperature ($T$) dependent Hall coefficient by solving Boltzmann transport equation in the two nesting cases, and compare with experiments. We find that, while experimental data in Y- and Hg-based cuprates\cite{HallY,HallHg} show a `continuous' sign reversal from negative to positive at a $T$ below the onset of pseudogap, the transition from an electron-pocket in the AN phase to large hole-FS in the paramagnetic state is discontinuous.\cite{HallY,HallHg} Finally, we write down a Ginzburg-Landau functional for the competing scenario between NN and AN phases, and propose a candidate phase diagram. An interesting manifestation of disorder pinned `chiral' CDW is also proposed.

\begin{figure}[top]
\rotatebox[origin=c]{0}{\includegraphics[width=0.9\columnwidth]{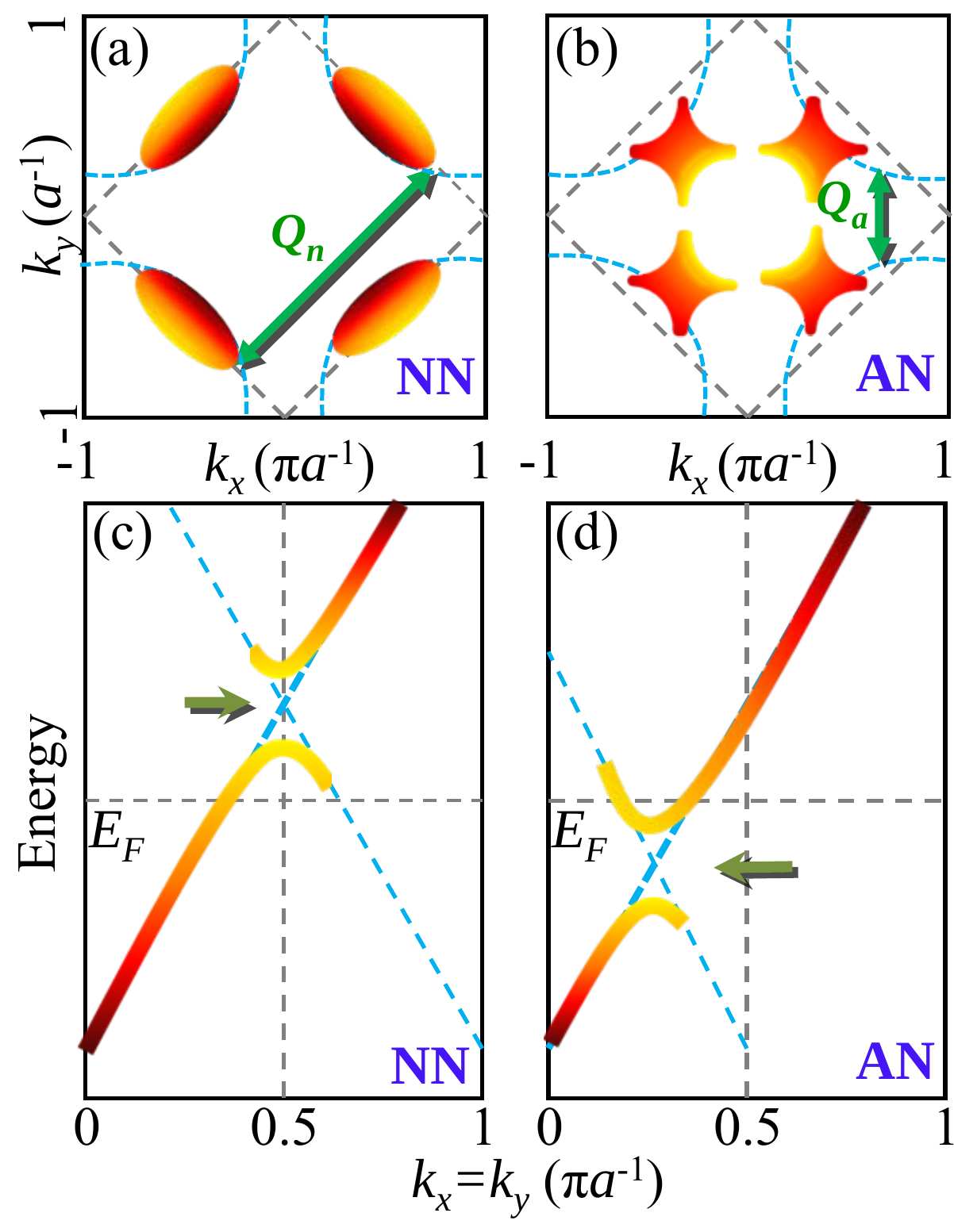}}
\caption{(Color online) (a) Schematic FS evolution for the NN at ${\bm Q}_n\rightarrow(\pi,\pi)$. (b) Same as (a) but for the AN at ${\bm Q}_a\rightarrow(\pm\pi/2,0),~(0,\pm\pi/2)$. (c)-(d) Electronic dispersion along the nodal direction for two cases discussed in their corresponding upper panels.}\label{fig1}
\end{figure}

In Fig.~\ref{fig1} we illustrate the NN and AN properties, and their differences in the electronic structure. In the NN phase, FSs across the magnetic zone boundary are nested, and thereby introduce a hole pocket centering at the nodal point as shown in Fig.~\ref{fig1}(a). The hole-pocket incipiently implies that the top of the lower split band crosses $E_F$, and a gap opens {\it in the empty state along the nodal direction}, see Fig.~\ref{fig1}(d). On the other hand, the biaxial AN nests the VHS regions of the FS, and thereby creates an electron-pocket whose center lies in between $\Gamma\rightarrow(\pi/2,\pi/2)$ and its equivalent directions as shown in Refs.~\onlinecite{npocket,BobCDW}, see Fig.~\ref{fig1}(b). The `nodal electron pocket' implies that the bottom of the upper split band lies below $E_F$, and {\it a gap opens in the filled state along the nodal direction} as illustrated in Fig.~\ref{fig1}(e).

\begin{figure}[top]
\rotatebox[origin=c]{0}{\includegraphics[width=.99\columnwidth]{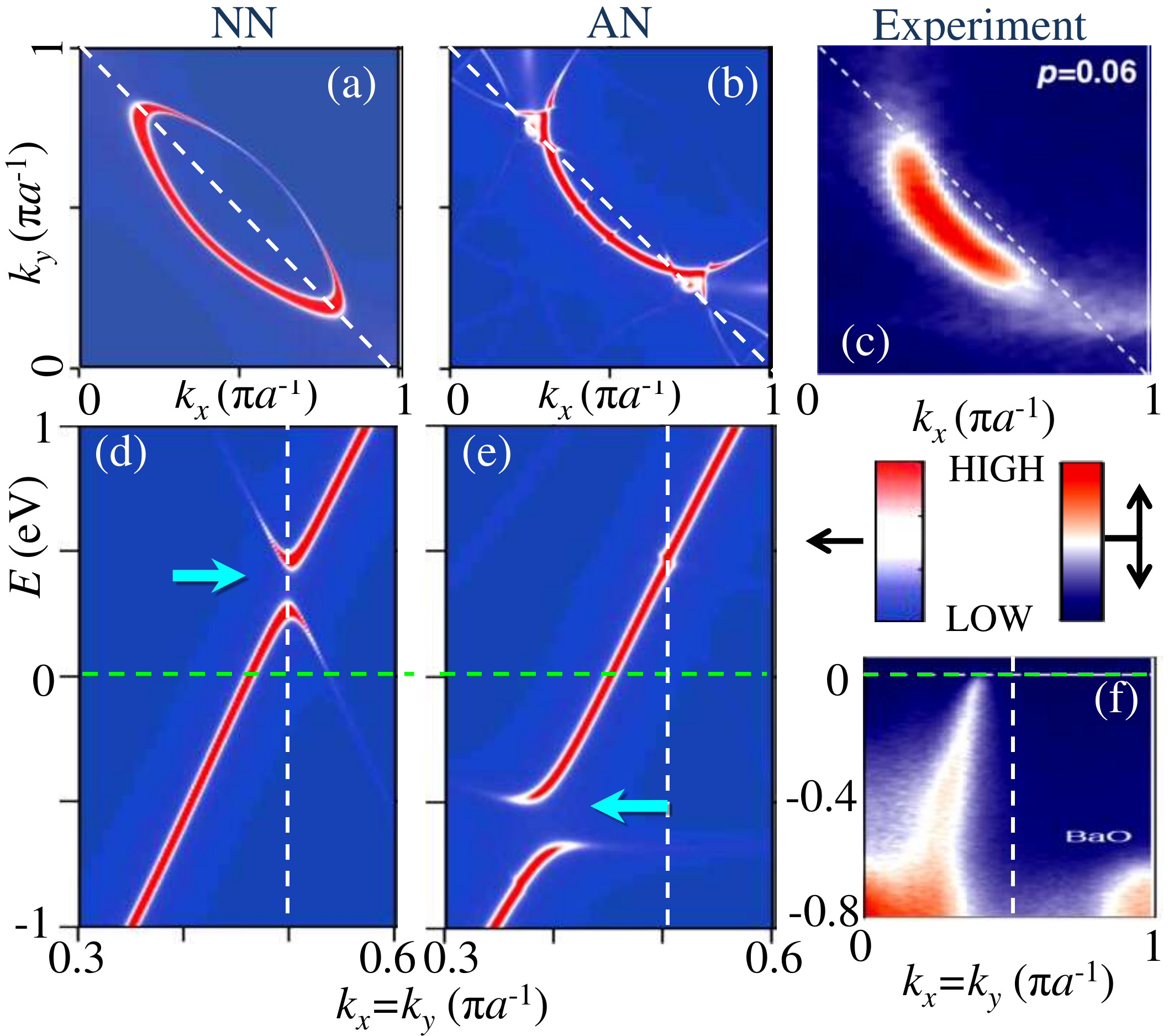}}
\caption{(Color online) (a) Computed FS for the NN at ${\bm Q}_n$. (b) Same as (a) but for the AN at ${\bm Q}_a$ (see text). (d)-(e) Computed dispersion along the nodal direction for the two cases discussed in their corresponding upper panels. (c)-(d) ARPES FS and dispersion along nodal line for underdoped YBCO$_{6.3}$.\cite{arpes}}\label{fig2}
\end{figure}

To provide a proof of principle, we perform a mean-field calculation using NN,\cite{Dastwogap} and AN,\cite{npocket} with same noninteracting starting point, and the corresponding results are shown in Fig.~\ref{fig2}. We use a one-band tight-binding model with parameters fitted to the {\it ab-initio} band-structure of YBa$_2$Cu$_3$O$_{6+x}$ (YBCO) given in Ref.~\cite{Dasresonance}. Using $\bm{Q}_n=(\pi,\pi)$, we obtain the quasiparticle spectral weight map at $E_F$ as shown in Fig.~\ref{fig2}(a), which gives the impression of the FS measured in ARPES.\cite{foot1} Using the same AN at $\bm{Q}_a^x=(\pi/2,0)$ and $\bm{Q}^y_a=(0,\pi/2)$ from Refs.~\cite{npocket} which presumably causes a CDW, we obtain the expected nodal electron-pocket as shown in Fig.~\ref{fig2}(b).
 The corresponding dispersion along the nodal direction is shown in Fig.~\ref{fig2}(d) which clearly reveals a gap opening below $E_F$. This is a robust result expected for any electron-pocket.

The ARPES FS, shown in Fig.~\ref{fig2}(c) for a representative case of underdoped YBCO$_{6.3}$, observes the main segment of the Fermi pocket or the so-called `Fermi arc'. ARPES FS can be considered to be consistent with both hole- or electron-pocket scenarios with the notion that it is difficult to detect the weak intensity of the shadow band which is present either on the front or on the back side of the main band, respectively. However, an important distinction between the hole- and electron-pockets along the nodal direction can be made via ARPES by searching for a gapless or gapped dispersion below $E_F$ along the nodal direction, respectively, as as shown in Fig.~\ref{fig2}(d)-(e).
The ARPES dispersion shown in Fig.~\ref{fig2}(f) does not reveal any such gap opening.

\begin{figure}
\rotatebox[origin=c]{0}{\includegraphics[width=.95\columnwidth]{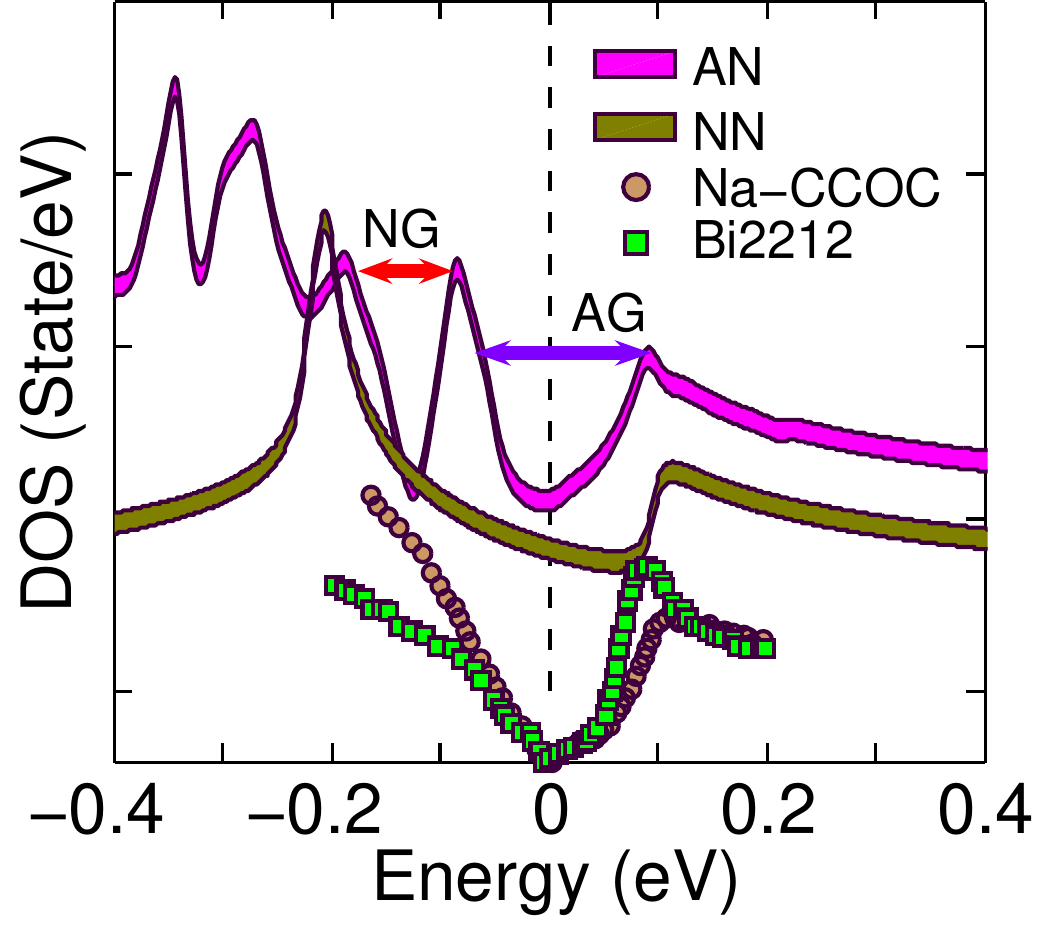}}
\caption{(Color online) Computed DOS for AN and NN cases (solid thick lines) are compared with STM results for two different hole-doped systems. The data for Ca$_{1.88}$Na$_{0.12}$CuO$_2$Cl$_2$ (Na-CCOC) and Bi$_2$Sr$_2$Dy$_{0.2}$Ca$_{0.8}$Cu$_2$O$_{8+\delta}$ (Bi2212) (normal state) are obtained from Ref.~\cite{STMdata}. The two horizontal arrows dictate the antinodal gap (AG) and nodal gap (NG) for the AN case.}\label{fig3}
\end{figure}

{\it DOS:-} The multiple gap structure for the AN, as compared to a single gap in the NN case is also evident in the density of states (DOSs), plotted in Fig.~\ref{fig3}. In both cases, the gap at the antinode (denoted as `AG') occurs at $E_F$ (dictated by purple horizontal arrow). For AN, the gap along the nodal axis (denoted as `NG') manifests as a separate gap in the DOS below $E_F$, marked by red horizontal arrow in Fig.~\ref{fig3}. For NN, however, the AG and NG (above $E_F$) are connected to each other via the `hot-spot' momenta, and thus appears as single gap. The STM results in the normal state for two hole-doped cuprates\cite{STMdata} (shown by different symbols), as available in this energy scale, do not show any signature of the second gap.

\begin{figure}
\rotatebox[origin=c]{0}{\includegraphics[width=.95\columnwidth]{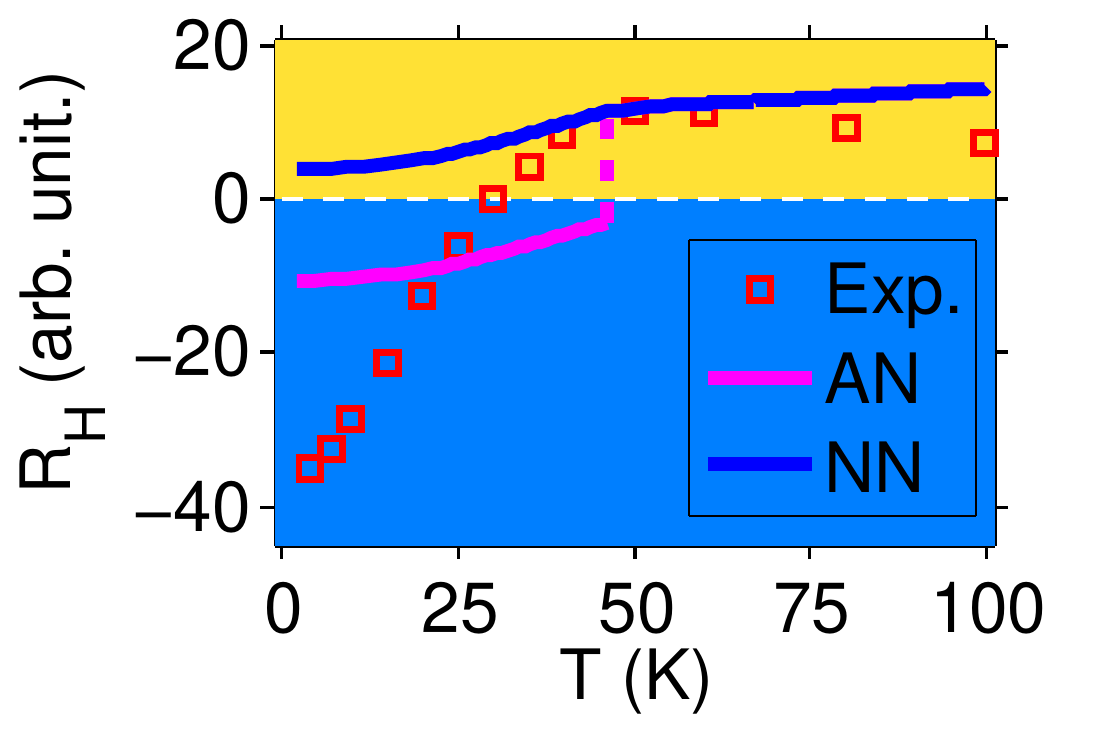}}
\caption{(Color online) Computed Hall coefficient, $R_H$ as a function of $T$, for the AN and NN cases. Symbols give experimental data for YBCO$_{6.51}$ at doping $x=0.1$ and magnetic field $B=55$~T, taken from Ref.~\onlinecite{HallY}. In both cases, the phase transition is assumed to occur at the same $T=55$~K. NN gives positive $R_H$ and connects smoothly to its paramagnetic value, whereas $R_H$ for the AN case is negative (coming from electron-like FS) below $T_a$, and at the transition, it shows a discontinuous jump (dashed line) to the positive value for the paramagnetic hole-like FS.}\label{fig4}
\end{figure}

{\it Hall effect:-} Hall coefficient, $R_H$, provides a crucial test of the nature of the quasiparticles on the FS, and its low-$T$ dependence gives valuable insights into the FS evolution, and the characteristic phase transition. Being interested in low-$T$, and low field, we employ a Boltzmann approach with momentum-independent quasiparticle scattering rate.\cite{Daschain} Furthermore, since our focus here is to compare the signatures of NN and AN on $R_H(T)$, we fix the same $T$-dependence of the gap to be of BCS-like as $\Delta(T)=\Delta_0(1-T/T_o)^{0.5}$, where $\Delta_0$ is the gap amplitude, taken to be same as in Fig.~\ref{fig2} and \ref{fig3}, and $T_o=55$~K is the same transition temperature. Sample results of $R_H(T)$ for NN and AN phase are shown in Fig.~\ref{fig4} which indeed reveal a sharp difference between them, both of which also depart from the experimental data.\cite{HallY} For AN, the electron-pocket ($R_H<0$) to paramagnetic hole-FS ($R_H>0$) transition at $T_o$ is discontinuous. For NN, although $R_H$ is smooth at the phase transition, a dominant negative $R_H$ is difficult to reproduce unless electron-like chain state in YBCO is taken into account\cite{Daschain}.
For HBCO, the negative $R_H$ indicates a phase imbalance between AN and NN,\cite{HallHg} and thus provides a unique opportunity to investigate a possible quantum tri-critical point, emerging from a coexistence between them.

{\em Ginzburg-Landau treatment:-} In this spirit, we study the stability of the two phases, and their possible coexistence at the level of Ginzburg-Landau functional argument. The Lagrangian of a system with competing interactions at ${\bm Q}_a$ and ${\bm Q}_n$ can be written in the Nambu decomposition of the  Grassmann (fermionic) field $\psi_{{\bm k},\sigma}$ as
%
\begin{eqnarray}
\mathcal{L}&=&\frac{1}{2}\sum_{{\bm k},\sigma,\omega_m}\left[\psi^{\dag}_{{\bm k},\sigma}G^{-1}_{\bm k}(i\omega_m)\psi_{k,\sigma} \right.\nonumber\\
&&~~~~~~~~+\sum_{i=a,n}\left\{\psi^{\dag}_{{\bm k}+{\bm Q}_i,\sigma}G^{-1}_{{\bm k}+{\bm Q}_i}(i\omega_m)\psi_{{\bm k}+{\bm Q}_i,\sigma}\right.\nonumber\\
&&~~~~~~~~~~~~~~~~\left.\left.+ U_i\psi^{\dag}_{{\bm k},\sigma}\psi_{k,\sigma}\psi^{\dag}_{{\bm k}+{\bm Q}_i,\sigma^{\prime}}\psi_{{\bm k}+{\bm Q}_i,^{\prime}}\right\}\right],
%
\label{Lagrangian}
\end{eqnarray}
where $\sigma$ denotes spin, and $\sigma^{\prime}$ is either the same spin for a CDW, or $d$-density wave or any phenomenological Umklapp process, or a spin flip for spin-ordering. The corresponding Green's functions are $G^{-1}({\bm k}^{\prime},\omega_n)=i\omega_n-\xi_{{\bm k}^{\prime}}$, for ${\bm k}^{\prime}={\bm k}, {\bm k}+{\bm Q}_{a/n}$, where $\omega_m$ is the Matsubara frequency and $\xi_{\bm k}$ is bare fermionic dispersion. The factor $1/2$ arises due to summing over the reduced Brillouin zone twice.

\begin{figure}[top]
\rotatebox[origin=c]{0}{\includegraphics[width=.99\columnwidth]{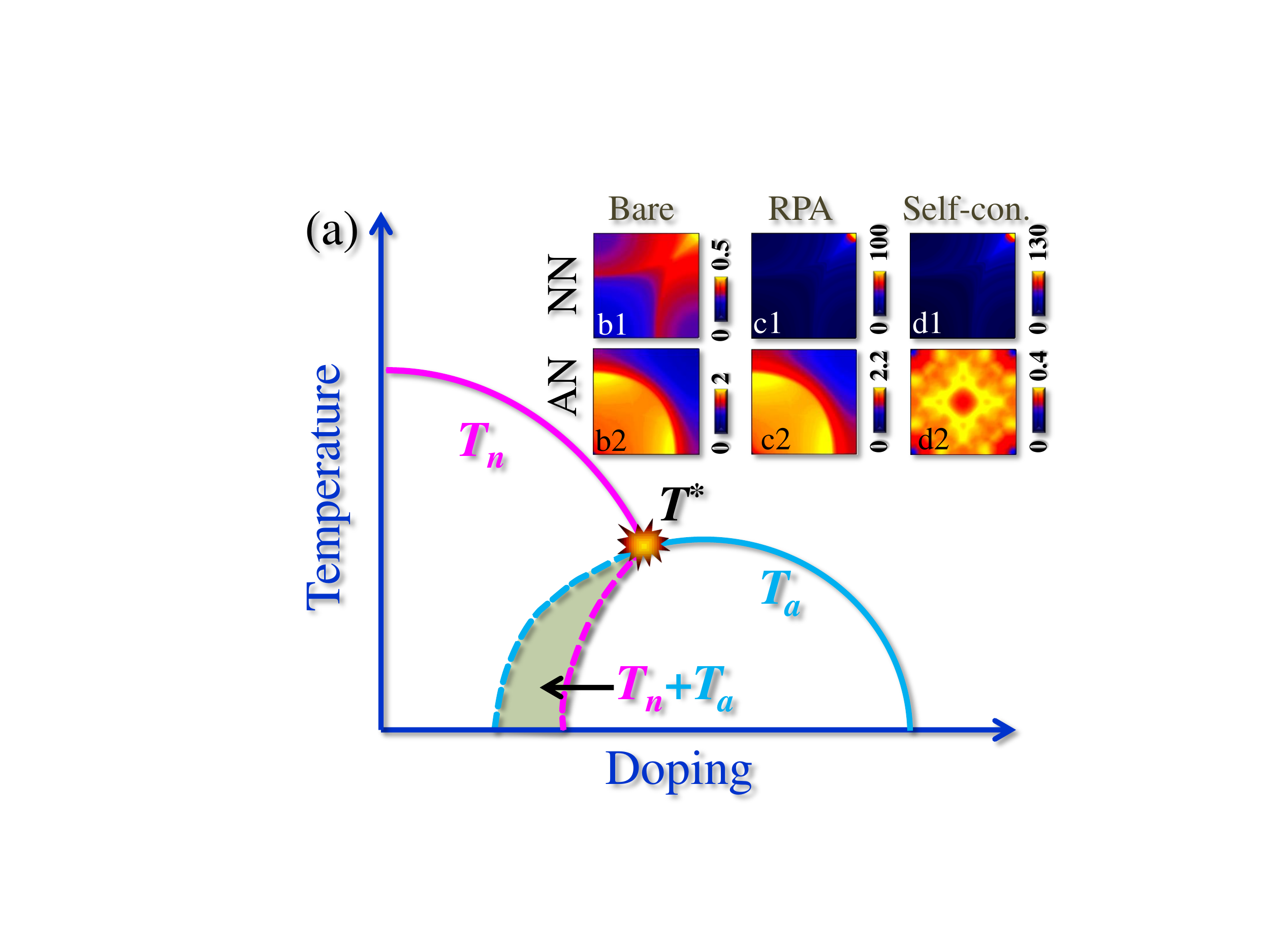}}
\caption{(Color online) (a) Phase diagram in ($x,T$)-plane for the AN ($T_a$) and NN ($T_n$) phases. The shaded area represents a possible phase coexistence region. $T^*$ is a common critical point of present interests. (b2)-(b2) The bare susceptibilities, plotted in two-dimensional momentum space at zero energy, show NN and AN at underdoped and optimally doped regions, respectively. (c1)-(c2) Corresponding RPA susceptibilities. (d1)-(d2) Self-consistent susceptibilittes in the corresponding gap states.}\label{fig5}
\end{figure}

We decouple the interaction terms into two corresponding bosonic fields $\Delta_{n/a}=U_{n/a}\sum_{{\bm k},s,t}\psi^{\dag}_{{\bm k}+{\bm Q}_n,s}[{\bm \sigma}/\delta]_{st}\psi_{{\bm k},t}$, by means of Hubbard-Stratanovich transformation, where ${\bm \sigma}$ gives the Pauli matrices. For the case of competing orders, the expansion of Eq.~\ref{Lagrangian} is standard,\cite{Rafael} which upto the quartic term of both fields (assuming they are real) becomes
\begin{eqnarray}
\mathcal{L}=\sum_{i=a,n}\left[\frac{\alpha_i}{2}(T-T_i)|\Delta_i|^2+\frac{\beta_i}{2}|\Delta_i|^4\right]+ \frac{\beta_{an}}{2}|\Delta_a|^2|\Delta_n|^2.\nonumber\\
\label{GL}
\end{eqnarray}
$T_{n/a}$ are the corresponding transition temperatures, and the expansion parameters $\alpha_i,~\beta_i$ are given in Ref.~\cite{expparam}. At the mean-field level, the leading instability for each order parameter stems from the logarithmic divergence of the corresponding susceptibility in the particle-hole channel. Since ${\bm Q}_a$ nests the antinodal region of the FS (see Fig.~\ref{fig5}), it is prone to reaching a singularity when the VHS approaches $E_F$ near or above the optimal doping, and drives the system to a CDW or ferromagnetic ordering.\cite{Bobchi} On the other hand, the NN, which leads to antiferromagnetism at half-filling dies off quickly with doping, see Fig.~\ref{fig5}(b1)-(b2), leaving a residual `hot-spot' instability at $Q_n$ with suppressed bare susceptibility in the two-dimensional system. The second order phase transitions of individual order can thus be monitored by these leading instability in the quadratic terms (Eq.~\ref{GL}).

Within the GL treatment, the competition and coexistence of two phases can be studied comprehensively near their common critical point at $T^*\approx T_a\approx T_n$.\cite{Rafael} A general formalism is obtained in the context of iron-pnictides that the Free energy for any competing orders of the form in Eq.~\ref{GL} drives a coexistence of the two order parameters if $\beta_a\beta_n-\beta_{an}^2>0$. $\beta_{a/n}$ correspond to the quartic Umklapp susceptibility with momentum transfer ${\bm Q}_{a/n}$, respectively, and a double Umklapp process involving both ${\bm Q}_a$ and ${\bm Q}_n$ generates the coupling term $\beta_{an}$. For the reasons given in the previous paragraph, near $T^*$, the non-interacting susceptibilities at ${\bm Q}_{n/a}$ governs $\beta_a>>\beta_n$, and $\beta_a>\beta_{an}>\beta_n$. To grasp qualitative insights, let us assume $\delta\geq0$ to be the same departure of $\beta_{a/n}$ from $\beta_{an}$ such that $\beta_{an}=\beta_a-\delta\approx\beta_n+\delta$, then the above condition for the coexistence reads $\delta^2-\delta(\beta_a-\beta_n)>0$. This implies that, for $\beta_a>\beta_n$, a phase coexistence is unfavored, and a first-order phase transition separates the AN and NN phases.

When many-body corrections are included in the Green's functions of the expansion parameters given in Ref.~\cite{expparam}, a second order phase transition can be monitored in two ways. Within a random-phase approximation (PRA), a strong divergence in the susceptibility can be obtained in the spin-channel at ${\bm Q}_n$, but not at ${\bm Q}_a$ below a critical value of $U$, see Fig.~\ref{fig5}(c1)-(c2). Furthermore, a self-consistent calculation makes the Green's function to be evaluated in the gapped quasiparticle state. Recalling results from Fig.~\ref{fig2}, both nestings gap out the antinodal region of the FS, and in turn, reduce the interacting susceptibility peak at ${\bm Q}_a$, see Fig.~\ref{fig5}(d1)-(d2). Both RPA and self-consistent scenarios thus promote $\beta_n\geq\beta_a$, driving a uniform phase coexistence, and hitherto a quantum tri-critical point at $T^*$ as shown in the phase diagram in Fig.~\ref{fig5}. Similar result was also proposed earlier in a different context.\cite{Rafael} The possibility of having a tri- or bi-critical point near the optimal doping clearly makes it an exciting problem for future study both experimentally and theoretically.

{\it Chiral charge oscillation:-} An interesting situation emerges when disorder pins one of the $\Delta_a^{(x/y)}$ domains only. This situation breaks in-plane rotational symmetry, as well as turns on a time-reversal breaking combination of $\Delta_a^{(x/y)}$  as $\Delta_a^t=\Delta_a^x\pm i\Delta_a^y$ with a finite expectation value of $\Delta_a^{t*}\Delta_a^t=|\Delta|^2$, where $\Delta$ is a real number. Rewriting $\Delta_a^t= |\Delta|e^{i\phi}$, we find that such scenario supports the presence of a Goldstone field $\phi$, according to Nambu-Goldstone theorem.\cite{NambuGoldstone} More interestingly, since the order parameter also breaks additional discrete crystal rotational symmetry, the emergent Goldstone mode becomes massive in this case. A $U(1)$ symmetry-induced current hence arises as ${\bm J}=-|\Delta|^2\partial_{\mu}\phi$, due to the spatial ($\mu=x,y$) variation of the order parameter around the disorder. The corresponding Lagrangian density that supplements to the total Free-energy functional in Eq.~\ref{GL} reads as
\begin{eqnarray}
\mathcal{L}^{\prime}&=&-\frac{1}{2}\left(\partial^{\mu}\Delta_a^{t*}\right)\left(\partial_{\mu}\Delta_a^{t}\right)+m^2\Delta_a^{t*}\Delta_a^t,\nonumber\\
&=&-\frac{|\Delta|^2}{2}\left(\partial^{\mu}\phi\right)\left(\partial_{\mu}\phi\right)+m^2|\Delta|^2.
\label{Goldstone}
\end{eqnarray}
Here the constant term $m$ has no physical significance to the Fermionic ensemble, since it merely shifts the overall energy scale. This special scenario gives an alternative explanation to the observations of both rotational,\cite{nematic,EAKim} and time-reversal symmetry breakings\cite{Campuzano,Kapitulnik,Greven} from solely charge ordering mechanism in doped systems, although other mechanisms to them exist.\cite{foot2,EAKim,Varma,Bansil,DasQ0}

Based on the present results, we conclude that the FS pocket or the segment of the FS observed in ARPES near nodal region is hole-like. Of course, such hole pocket scenario cannot explain the electron-like FS predicted by numerous magneto-resistance measurements. For the NN, electron-like FSs appear near the antinodal region close to the quantum-(bi- or tri-) critical point of the pseudogap where its strength is weak. Since such electron-pocket appears in the region where the FS is in the verge of becoming the large metallic FS, it is difficult to experimentally separate out the presence of electron-pocket.\cite{DasEP} For YBCO, however, the chain state is electron-like and contribute to its large negative Hall coefficient.\cite{DasEP} Our obtained results suggest that the  CDW modulation is preferably a secondary order, which is either phase separated or coexists in a narrow doping range with the NN order.

\begin{acknowledgments}
The author thanks P. Werner for the encouragement to write up this work, and is also indebted to H. Alloul, R. S. Markiewicz, A. Balatsky, M. Vojta, and N. Harrison for numerous stimulating discussions. The work is supported by the U.S. DOE through the Office of Science (BES) and the LDRD Program and facilitated by NERSC computing allocation.
\end{acknowledgments}


\begin{thebibliography}{10}
\bibitem{foot1} We do not make any specific assumption of the particular nature of order parameter that can emerge in this case, as the macroscopic gap opening and the FS properties do not rely on its microscopic origin.\cite{Dastwogap}
\bibitem{expparam} The expansion parameters are
$\alpha_{a/n}\Rightarrow \langle G_{\bm k}(i\omega_m)G_{\bm k}(i\omega_m)+G_{{\bm k}+{\bm Q}_{a/n}}(i\omega_m)G_{{\bm k}+{\bm Q}_{a/n}}(i\omega_m)\rangle$, 
%
$\beta_{a/n}\Rightarrow 2T\langle G^2_{\bm k}(i\omega_m)G^2_{{\bm k}+{\bm Q}_{a/n}}(i\omega_m)\rangle,$ 
%
$\beta_{an} \Rightarrow 4T\langle G^2_{\bm k}(i\omega_m)G_{{\bm k}+{\bm Q}_{a}}(i\omega_m)G_{{\bm k}+{\bm Q}_{n}}(i\omega_m)\rangle$, 
%
where $\langle...\rangle\rightarrow\int\frac{d^2k}{(2\pi)^2}\sum_{\omega_m}$.
\end{thebibliography}
\end{document}